\documentclass[10pt,conference]{IEEEtran}
\IEEEoverridecommandlockouts

\usepackage{cite}
\usepackage{amsmath,amssymb,amsfonts}
\usepackage{textcomp}
\usepackage{xcolor}
\usepackage{booktabs}
\usepackage{amsmath}
\usepackage{pgfplots}
\pgfplotsset{compat=1.18}
\usetikzlibrary{patterns}
\usepackage{placeins}
\usepackage{stfloats}
\usepackage{booktabs}
\usepackage{multirow}
\usepackage{tikz}
\usepackage{pgfplots}
\pgfplotsset{compat=1.17}
\usepackage{amsmath,amssymb}
\usepackage{graphicx}
\usepackage{booktabs}
\usepackage{multirow}
\usepackage{algorithm}
\usepackage{algorithmic}
\usepackage{needspace}
\begin{document}

\title{SLA-Safe Energy Control for AI-Native NG-RAN Using Stability-Aware Constrained PPO}

\author{
\IEEEauthorblockN{Dharmendra Kumar}
\IEEEauthorblockA{
Independent Researcher\\
Dallas, TX, USA\\
dharmendra.sharma4@gmail.com
}
\thanks{\textcopyright{} 2026 IEEE. Personal use of this material is
permitted. Permission from IEEE must be obtained for all other uses.
Accepted for publication in the Proceedings of the 2026 IEEE PIMRC
Workshops, Singapore.}
}

\maketitle

\begin{abstract}
Artificial intelligence is becoming a key enabler for AI-native radio access networks, where closed-loop learning-based controllers can support adaptive and autonomous network optimization. One important AI-for-RAN use case is energy saving, in which radio resources and cell energy modes must be dynamically controlled without violating user quality-of-service (QoS) or service-level agreement (SLA) requirements. However, aggressive sleep-state or deactivation decisions may reduce energy consumption at the cost of throughput degradation, delay increase, SLA violations, and unstable mode switching, especially under time-varying and bursty traffic conditions. This paper proposes a stability-aware constrained reinforcement learning framework for SLA-safe energy control in 5G NG-RAN. The problem is formulated as a constrained Markov decision process in which an AI-native controller selects closed-loop energy-saving actions based on cell load, queue status, active-user information, current energy mode, and SLA-related indicators. The proposed framework uses constrained proximal policy optimization with adaptive Lagrangian penalties to account for throughput, delay, and SLA constraints. To improve operation under traffic distribution shift, the controller is trained using mixed nominal and stress traffic regimes, while a switching-stability penalty is introduced to reduce oscillatory transitions between active and low-power modes. Simulation results in a seven-cell NG-RAN environment show that the proposed controller reduces energy consumption by approximately 41.4\% under nominal traffic, 10.5\% under stress traffic, and 22.9\% under unseen-stress traffic relative to the Always-On baseline. Under stress and unseen-stress traffic, the controller preserves zero SLA violation and zero throughput loss, indicating service-preserving operation under challenging conditions. The proposed method also reduces switching activity compared with basic threshold-based energy saving. These results show that stability-aware constrained reinforcement learning can provide a balanced AI-native RAN control approach for energy efficiency, SLA safety, robustness under traffic shift, and operational stability.
\end{abstract}


\section{Introduction}

Future radio access networks are expected to evolve from rule-based optimization toward AI-native closed-loop control, where learning-based agents observe network conditions, make autonomous decisions, and adapt to changing traffic and service requirements. Among AI-for-RAN use cases, network energy saving is especially relevant because radio access infrastructure contributes significantly to mobile network power consumption, and energy-saving actions must be applied without degrading quality of service (QoS) or service-level agreement (SLA) compliance.

Energy-saving control in NG-RAN is challenging because it is not only a power-minimization problem. Actions such as cell sleep, carrier deactivation, or transitions to low-power modes can reduce energy consumption during light-load periods, but aggressive decisions may degrade throughput, increase delay, and trigger SLA violations when traffic suddenly increases, and frequent mode transitions may further cause unstable control behavior. These challenges are amplified by traffic nonstationarity, where a policy trained under nominal conditions may fail under bursty or stress traffic.

Traditional threshold-based energy-saving methods are simple and interpretable, but they often depend on fixed load thresholds and may not adapt well to changing traffic patterns. Reinforcement learning can learn adaptive control policies from network feedback; however, reward-only learning may still select actions that reduce energy while violating delay, throughput, or SLA requirements. Therefore, AI-native RAN energy control requires a constrained closed-loop formulation in which QoS and SLA requirements are treated as explicit operational constraints, while switching stability is also considered.

In this paper, we propose a stability-aware constrained reinforcement learning framework for SLA-safe energy control in AI-native NG-RAN. The proposed framework formulates network energy saving as a constrained Markov decision process. An AI-RAN controller observes RAN telemetry such as cell load, queue status, active-user information, current energy mode, and SLA-related indicators, and selects energy-saving actions such as active, light-sleep, or deep-sleep modes. The controller is trained using constrained proximal policy optimization with adaptive Lagrangian penalties for throughput, delay, and SLA constraints. To improve robustness under traffic distribution shift, training incorporates both nominal and stress traffic regimes. A switching-stability penalty is also introduced to reduce oscillatory transitions between energy modes.

The main contributions of this paper are summarized as follows:
\begin{itemize}
\item We formulate AI-native NG-RAN energy saving as a constrained closed-loop control problem that jointly considers energy efficiency, throughput, delay, SLA compliance, and switching stability.

\item We develop a stability-aware constrained PPO framework with adaptive Lagrangian penalties, mixed-regime training, and switching-cost regularization for SLA-safe energy control.

\item We evaluate the proposed approach in a seven-cell NG-RAN simulation under nominal, stress, and unseen-stress traffic scenarios, showing energy reductions of approximately 41.4\%, 10.5\%, and 22.9\% relative to Always-On, respectively.

\item We show that constrained learning and mixed-regime training are important for SLA-safe operation, while switching-aware control reduces unnecessary mode transitions compared with basic threshold-based energy saving.
\end{itemize}

\section{Related Work and Gap}

AI/ML techniques are increasingly considered for data-driven, closed-loop RAN optimization, including energy saving, load balancing, mobility optimization, and interference coordination. This work belongs to the AI-for-RAN category, where AI is used to improve RAN operation rather than to host AI workloads on RAN infrastructure~\cite{b1,b2,b4,b5}.

Energy saving in cellular networks has traditionally been addressed using rule-based or threshold-based mechanisms such as cell sleep, carrier shutdown, and low-power mode selection. These approaches are simple and interpretable, but their performance depends on manually selected thresholds and assumptions about traffic behavior. Under dynamic or bursty traffic, fixed thresholds may either miss energy-saving opportunities or select aggressive low-power actions that degrade throughput, increase delay, or violate SLA requirements. Frequent transitions between active and low-power modes can also create unstable control behavior.

Reinforcement learning has been widely studied for wireless and RAN optimization because it can learn adaptive policies from network feedback~\cite{b6,b10,b11,b12}. However, reward-only RL formulations may reduce average energy consumption while still producing unsafe behavior under stress traffic. Safe and constrained reinforcement learning provides a more suitable formulation for RAN control because QoS and SLA requirements can be represented as explicit constraints~\cite{b8}. Nevertheless, applying constrained RL to AI-native RAN energy saving remains challenging because the policy must also handle traffic distribution shift and avoid unstable mode switching.

Recent learning-based energy-saving studies also address switching behavior and realistic RAN evaluation: Bordin \textit{et al.}~\cite{b13} train PPO/DQN agents with a switching-cost term to reduce cell activation ping-ponging in an ns-O-RAN environment, while Bassoy \textit{et al.}~\cite{b14} propose a deep-RL energy-efficiency scheme for 6G RAN. Our framework additionally treats SLA, delay, and throughput as explicit constraints via adaptive Lagrangian penalties rather than reward terms alone, combined with mixed-regime training for SLA-safe operation under both nominal and stress conditions.

The gap addressed in this paper is the lack of a unified closed-loop framework that jointly considers energy efficiency, QoS preservation, SLA compliance, traffic-shift robustness, and switching stability. To address this gap, we propose a stability-aware constrained PPO framework for SLA-safe energy control in AI-native NG-RAN. The proposed method combines adaptive Lagrangian constraint handling, mixed-regime training across nominal and stress traffic, and switching-cost regularization to learn an energy-saving policy that remains service-aware under changing traffic conditions.

\section{AI-Native RAN Control Architecture}

This section presents the AI-native closed-loop RAN control architecture considered in this work. The objective is to support SLA-safe network energy saving by allowing an AI-RAN controller to observe RAN telemetry, select energy-saving actions, and receive feedback on energy consumption, QoS performance, SLA compliance, and control stability. Energy saving is treated as an AI-for-RAN use case, where intelligence is used to optimize RAN operation rather than to host AI workloads on the RAN infrastructure~\cite{b1,b3,b4,b5}.

\begin{figure*}[t]
\centering
\includegraphics[width=0.82\textwidth]{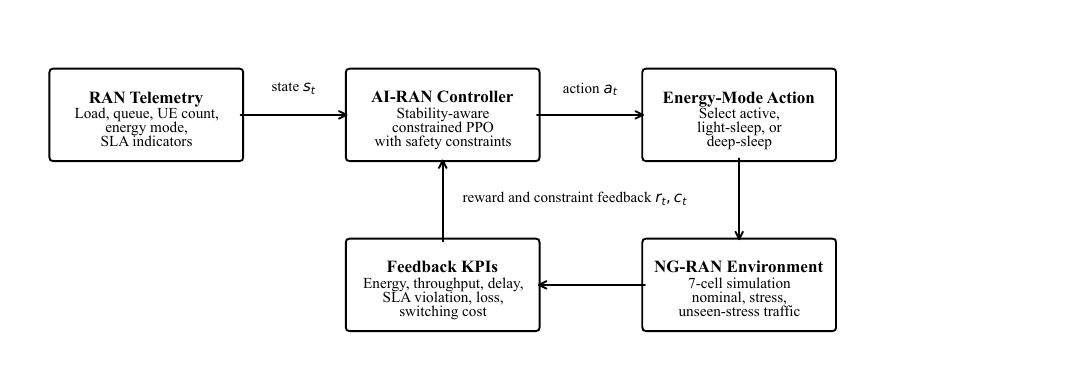}
\caption{AI-native closed-loop RAN energy-control architecture. The controller observes RAN telemetry as state $s_t$, selects an energy-control action $a_t$, and receives reward and constraint feedback based on energy consumption, QoS performance, SLA compliance, throughput loss, and switching stability.}
\label{fig}
\end{figure*}

Fig.~\ref{fig} illustrates the proposed AI-native closed-loop RAN energy-control architecture. The controller observes RAN telemetry as state $s_t$, selects an energy-control action $a_t$, and receives reward and constraint feedback based on energy consumption, QoS performance, SLA compliance, throughput loss, and switching stability.

\subsection{Closed-Loop Control View}

At each decision epoch, the RAN environment provides telemetry information to the AI-RAN controller, including cell load, queue status, active-user information, current energy mode, and recent QoS or SLA indicators. Based on this state, the controller selects an energy-mode action for each controlled cell, such as active, light-sleep, or deep-sleep operation. After the action is applied, the environment returns feedback in terms of energy consumption, throughput, delay, SLA violation, throughput loss, and switching activity.

This closed-loop interaction allows the controller to learn the long-term effect of energy-saving actions: a low-power decision may reduce energy immediately but increase delay or SLA violation risk if traffic demand increases shortly afterward, so the controller must learn not only when to save energy, but also when to preserve resources for service reliability.

The architecture combines three design elements. First, telemetry-driven state observation enables adaptation to current RAN conditions. Second, constrained learning models throughput, delay, and SLA requirements as explicit operational constraints rather than only as soft reward penalties. Third, switching-aware regularization discourages unnecessary transitions between active and low-power modes, improving closed-loop stability. 

To address traffic distribution shift, the proposed framework uses mixed-regime training with both nominal and stress traffic episodes, enabling energy reduction under regular load while adapting toward service-preserving operation under stress conditions.

\section{Problem Formulation}

This section formulates AI-native NG-RAN energy saving as a constrained closed-loop decision-making problem. The objective is to reduce long-term RAN energy consumption while preserving QoS, maintaining SLA compliance, and avoiding unstable switching between energy modes.

\subsection{System Model}

We consider a multi-cell NG-RAN system with $B$ controllable cells. Time is divided into discrete decision epochs indexed by $t$. At each epoch, the AI-RAN controller observes the current network condition and selects an energy mode for each cell. Let $m_b(t)$ denote the energy mode of cell $b$ at time $t$, where $b \in \{1,\ldots,B\}$. The total network energy consumption is

\begin{equation}
E_{\mathrm{tot}}(t)
=
\sum_{b=1}^{B}
E_b\big(m_b(t),L_b(t)\big)
\end{equation}

where $E_b(\cdot)$ denotes the energy consumed by cell $b$ under mode $m_b(t)$ and load $L_b(t)$. Lower-power modes reduce energy consumption but may also reduce service capacity and increase the risk of QoS degradation.

\subsection{State and Action Spaces}

At each decision epoch, the controller observes the state vector

\begin{equation}
s_t =
[L(t),Q(t),N(t),M(t),\Gamma(t)]
\end{equation}

where $L(t)$ is the cell-load vector, $Q(t)$ represents queue-related indicators, $N(t)$ is the active-UE vector, $M(t)$ denotes the current energy-mode vector, and $\Gamma(t)$ contains recent QoS or SLA-related indicators.

The action selected by the controller is

\begin{equation}
a_t =
[m_1(t),m_2(t),\ldots,m_B(t)]
\end{equation}

where each cell mode belongs to the discrete action set

\begin{equation}
m_b(t)
\in
\{\text{active},\text{light-sleep},\text{deep-sleep}\}.
\end{equation}

\subsection{Objective and Service Constraints}

A purely energy-minimizing policy may select aggressive low-power actions that reduce energy but degrade throughput, increase delay, or violate SLA targets. Therefore, the control problem includes explicit QoS and SLA constraints.

Let $\pi(a_t|s_t)$ denote the control policy over a decision horizon $T$. The long-term energy objective is

\begin{equation}
\begin{aligned}
J_E(\pi)
&=
\mathbb{E}_{\pi}
\left[
\sum_{t=0}^{T-1}
E_{\mathrm{tot}}(t)
\right], \\
\pi^{\star}
&=
\arg\min_{\pi}
J_E(\pi).
\end{aligned}
\label{eq:energy_objective}
\end{equation}

Let $R(t)$ denote achieved throughput, $D(t)$ denote delay, and $V_{\mathrm{sla}}(t)$ denote the SLA violation indicator or violation rate. The service constraints are

\begin{equation}
\begin{aligned}
\mathbb{E}_{\pi}[V_{\mathrm{sla}}(t)]
&\leq \epsilon_{\mathrm{sla}}, \\
\mathbb{E}_{\pi}[D(t)]
&\leq \epsilon_d, \\
\mathbb{E}_{\pi}[R(t)]
&\geq \epsilon_r,
\end{aligned}
\end{equation}

where $\epsilon_{\mathrm{sla}}$, $\epsilon_d$, and $\epsilon_r$ are predefined SLA, delay, and throughput thresholds.

\subsection{Reward, Constraint Costs, and CMDP}

The reward balances energy efficiency, service utility, and switching stability:

\begin{equation}
\begin{aligned}
r_t
&= -\alpha E_{\mathrm{tot}}(t)
+ \beta U_{\mathrm{serv}}(t) \\
&\quad - \gamma C_{\mathrm{sw}}(t)
\end{aligned}
\end{equation}

where $U_{\mathrm{serv}}(t)$ captures service utility based on throughput and delay, $C_{\mathrm{sw}}(t)$ is the switching cost, and $\alpha$, $\beta$, and $\gamma$ are nonnegative weighting coefficients. The switching cost is

\begin{equation}
C_{\mathrm{sw}}(t)
=\sum_{b=1}^{B}
\mathbb{I}_{b}^{\mathrm{sw}}(t)
\end{equation}

where $\mathbb{I}_{b}^{\mathrm{sw}}(t)=1$ if $m_b(t)\neq m_b(t-1)$ and $0$ otherwise.

\Needspace{10\baselineskip}
The constraint costs are defined as
\begin{equation}
\begin{aligned}
c_1(t) &= V_{\mathrm{sla}}(t), \\
c_2(t) &= \left[D(t)-\epsilon_d\right]^+, \\
c_3(t) &= \left[\epsilon_r-R(t)\right]^+ .
\end{aligned}
\label{eq:constraint_costs}
\end{equation}

The reward and constraint returns are defined as
\begin{equation}
\begin{aligned}
J_r(\pi)
&=
\mathbb{E}_{\pi}
\left[
\sum_{t=0}^{T-1} r_t
\right], \\
J_{c_i}(\pi)
&=
\mathbb{E}_{\pi}
\left[
\sum_{t=0}^{T-1} c_i(s_t,a_t)
\right],
\quad i=1,2,3 .
\end{aligned}
\label{eq:returns}
\end{equation}
where $c_i(s_t,a_t)$ denotes the instantaneous realization of cost $c_i(t)$ under state $s_t$ and action $a_t$.

The constrained optimization problem is formulated as
\begin{equation}
\begin{aligned}
\max_{\pi}\;& J_r(\pi) \\
\mathrm{s.t.}\;& J_{c_i}(\pi)\leq \zeta_i,
\quad i=1,2,3,
\end{aligned}
\label{eq:cmdp_objective}
\end{equation}
where $\zeta_i$ denotes the allowable cumulative constraint budget for constraint $i$.

Using Lagrangian relaxation, the constrained objective becomes
\begin{equation}
\begin{aligned}
\mathcal{L}(\pi,\lambda)
&=
J_r(\pi)
-
\sum_{i=1}^{3}
\lambda_i
\left(
J_{c_i}(\pi)-\zeta_i
\right),
\end{aligned}
\label{eq:lagrangian}
\end{equation}
where $\lambda_i \geq 0$ are Lagrange multipliers that penalize constraint-budget violations during training~\cite{b8}.

\section{Stability-Aware Constrained PPO Framework}

This section presents the proposed stability-aware constrained PPO framework for SLA-safe energy control in AI-native NG-RAN. The framework combines PPO-based policy learning, adaptive Lagrangian constraint handling, mixed-regime training, and switching-aware regularization.

\subsection{Framework Overview}

At each decision epoch, the controller observes the RAN state $s_t$ and selects an energy-saving action $a_t$ according to a policy $\pi_{\theta}(a_t|s_t)$ parameterized by $\theta$. The selected action determines the energy mode of each controlled cell. After the action is applied, the environment returns the reward $r_t$, constraint costs $c_i(t)$, switching cost $C_{\mathrm{sw}}(t)$, and next state $s_{t+1}$. The policy is updated using trajectories collected from closed-loop interaction with the simulated RAN environment.

The proposed framework differs from reward-only RL in two main ways. First, QoS and SLA requirements are handled as explicit constraints using adaptive Lagrangian penalties. Second, switching stability is incorporated into the reward design to discourage unnecessary transitions between active and low-power modes.

\subsection{Constrained PPO Update}

Let $L_{\mathrm{PPO}}(\theta)$ denote the standard PPO clipped surrogate objective~\cite{b7}. The constrained PPO objective is written in compact form as

\begin{equation}
\begin{aligned}
\mathcal{J}(\theta)
&=
L_{\mathrm{PPO}}(\theta)
-\sum_{i=1}^{3}
\lambda_i
\left(
\hat{J}_{c_i}(\theta)-\zeta_i
\right)
\end{aligned}
\end{equation}

where $\hat{J}_{c_i}(\theta)$ is the empirical cumulative cost for constraint $i$, $\zeta_i$ is the corresponding constraint budget, and $\lambda_i$ is the Lagrange multiplier. The multipliers are updated as

\begin{equation}
\begin{aligned}
\lambda_i
&\leftarrow
\left[
\lambda_i+
\eta_{\lambda}
\left(
\hat{J}_{c_i}(\theta)-\zeta_i
\right)
\right]^{+},
\end{aligned}
\end{equation}

where $[x]^{+}=\max\{0,x\}$ and $\eta_{\lambda}$ is the multiplier learning rate. This update increases the penalty when constraint costs exceed the allowable budget and reduces the effective pressure when the policy remains within feasible limits.

\subsection{Switching Stability and Mixed-Regime Training}

Frequent transitions between active and low-power modes may create oscillatory behavior and service fluctuation. To improve control stability, the reward includes the switching penalty $C_{\mathrm{sw}}(t)$ defined in Section IV, which discourages unnecessary mode changes and encourages smoother control behavior near mode-selection boundaries.

Traffic in operational RAN systems is nonstationary, and a policy trained only under nominal traffic may learn an aggressive low-energy strategy that performs poorly under bursty or stress traffic. To improve robustness under traffic distribution shift, the proposed framework uses mixed-regime training, sampling episodes from both nominal and stress traffic regimes: nominal episodes expose the controller to regular load variations and energy-saving opportunities, while stress episodes expose it to sudden demand increases and force it to learn when energy saving should be reduced to preserve QoS and SLA performance.

During evaluation, the trained policy operates in closed loop without further parameter updates. It is evaluated under nominal, stress, and unseen-stress traffic scenarios using energy consumption, throughput, delay, SLA violation rate, throughput loss, and switching activity.

\section{Experimental Setup}

\subsection{Simulation Environment and Traffic Scenarios}

The proposed framework is evaluated in a controlled multi-cell NG-RAN simulation environment. The main evaluation uses a seven-cell layout consisting of one center cell and six neighboring cells. Each cell can operate in one of three energy modes: active, light-sleep, and deep-sleep. Time is divided into discrete decision epochs, and each decision epoch represents one control interval of 1~s. At each epoch, the AI-RAN controller observes the network state, selects an energy mode for each controlled cell, and receives feedback from the environment.

\begin{table}[t]
\caption{Main Simulation Parameters}
\label{tab:sim_params}
\centering
\footnotesize
\setlength{\tabcolsep}{3pt}
\renewcommand{\arraystretch}{0.95}
\begin{tabular}{@{}p{0.38\columnwidth}p{0.54\columnwidth}@{}}
\hline
\textbf{Parameter} & \textbf{Value} \tabularnewline
\hline
Network layout & 7 cells \tabularnewline
Energy modes & Active, light-sleep, deep-sleep \tabularnewline
Control interval & 1 s \tabularnewline
Episode length & 500 decision epochs \tabularnewline
Nominal load & 10--25 UEs per cell \tabularnewline
Stress load & 35--60 UEs per cell \tabularnewline
Evaluation seeds & 5 random seeds \tabularnewline
Evaluation scenarios & Nominal, stress, unseen-stress \tabularnewline
\hline
\end{tabular}
\end{table}

Three traffic scenarios are considered. The nominal scenario represents regular traffic variation with moderate load fluctuations, where the number of active UEs varies between 10 and 25 per cell. The stress scenario represents bursty demand and sudden load increases, where the number of active UEs varies between 35 and 60 per cell with random burst periods. The unseen-stress scenario is used only during evaluation and is not included in training. It uses different burst intensity, duration, or hotspot location compared with the stress traffic used during training. During mixed-regime training, each episode is sampled from either the nominal or stress traffic regime, while unseen-stress is reserved for testing only.

\subsection{Energy, Service, and Baseline Models}

A normalized energy model represents the relative energy cost and service capacity of different cell modes. Active, light-sleep, and deep-sleep modes use normalized energy costs of 1.00, 0.60, and 0.30, with corresponding capacity factors of 1.00, 0.70, and 0.35, respectively. Deep-sleep provides the largest energy reduction but also reduces available service capacity, increasing the risk of delay growth or throughput degradation under high load.

QoS performance is evaluated using throughput and delay, while SLA compliance is evaluated using an SLA violation rate. In this work, an SLA violation occurs when the delay exceeds the predefined delay threshold or when throughput loss relative to the Always-On baseline exceeds the predefined throughput-loss threshold. Throughput loss is computed as

\begin{equation}
R_{\mathrm{loss}}(t)
=\frac{R_{\mathrm{AO}}(t)-R(t)}
{R_{\mathrm{AO}}(t)}
\end{equation}

where $R_{\mathrm{AO}}(t)$ is the throughput achieved by the Always-On baseline and $R(t)$ is the throughput achieved by the evaluated method. Switching activity is measured using $C_{\mathrm{sw}}(t)$.

The proposed method is compared against rule-based baselines (Always-On, Threshold-ES, and Threshold-ES with hysteresis, which uses separate activation/deactivation thresholds to reduce frequent switching) and learning-based ablations: Unconstrained PPO (no explicit SLA, delay, or throughput constraints), Constrained PPO without switching penalty, and Nominal-Only PPO (trained only under nominal traffic) to assess the importance of mixed-regime training.

\subsection{Training Configuration and Evaluation Metrics}

The proposed controller is trained using PPO with an actor-critic neural network architecture. The policy and value networks are multilayer perceptrons with two hidden layers of 64 units each. Unless otherwise stated, the learning rate is $3\times10^{-4}$, discount factor is $0.99$, GAE parameter is $0.95$, PPO clip range is $0.20$, batch size is 64, rollout length is 1024 steps, entropy coefficient is $0.01$, and Lagrange multiplier learning rate is $1\times10^{-3}$. The same network architecture and training budget are used for the learning-based baselines.

All methods are evaluated using normalized energy consumption, aggregate throughput, average delay, SLA violation rate, throughput loss relative to Always-On, and switching activity. Results are averaged over five random seeds and reported using mean values. The proposed method is considered effective if it reduces energy consumption under nominal traffic, maintains near-Always-On QoS under stress and unseen-stress traffic, and avoids excessive switching activity.

\subsection{Computational Considerations}

Training the constrained PPO agent for 200,000 timesteps takes approximately 14 minutes on a single CPU core. The lightweight two-layer, 64-unit MLP actor requires only a single forward pass per decision epoch, with measured mean inference latency of 0.74~ms per cell (2000 trials, CPU)---negligible relative to the 1~s control interval and suitable for real-time RAN deployment.

\section{Results and Discussion}

This section evaluates the proposed stability-aware constrained PPO framework under nominal, stress, and unseen-stress traffic scenarios. The evaluation focuses on the tradeoff among energy consumption, QoS preservation, SLA compliance, throughput loss, and switching activity.

\subsection{Overall Performance}

Table~\ref{tab:main_results} summarizes the comparison between the proposed method and rule-based baselines. Under nominal traffic, the proposed CRL-PPO controller reduces normalized energy consumption from 0.8567 to 0.5016, corresponding to approximately 41.4\% energy saving relative to Always-On, while maintaining low delay, small throughput loss, and an SLA violation rate below 1\%. Under stress traffic, the controller increases resource activation and operates closer to Always-On, achieving 10.5\% energy saving with zero SLA violation and zero throughput loss. In the unseen-stress scenario, which is not included in training, the proposed method achieves 22.9\% energy saving while maintaining zero SLA violation, zero throughput loss, and delay equal to the Always-On reference.

\begin{table*}[t]
\centering
\caption{Main Performance Comparison Across Traffic Scenarios}
\label{tab:main_results}
\scriptsize
\setlength{\tabcolsep}{4pt}
\renewcommand{\arraystretch}{0.95}
\begin{tabular*}{\textwidth}{@{\extracolsep{\fill}}llcccccc@{}}
\hline
\textbf{Scenario} & \textbf{Method} & \textbf{Energy} & \textbf{Throughput} & \textbf{Delay} & \textbf{SLA Viol.} & \textbf{Thr. Loss} & \textbf{Switch Rate} \tabularnewline
\hline
Nominal & Always-On & 0.8567 & 0.2836 & 0.0010 & 0.0000 & 0.0000 & 0.0000 \tabularnewline
Nominal & Threshold-ES & 0.3686 & 0.2836 & 0.0010 & 0.0000 & 0.0000 & 0.1605 \tabularnewline
Nominal & Threshold-ES-Hysteresis & 0.3550 & 0.2837 & 0.0010 & 0.0008 & 0.0008 & 0.0089 \tabularnewline
Nominal & \textbf{Proposed-CRL-PPO} & 0.5016 & 0.2840 & 0.0011 & 0.0072 & 0.0025 & 0.0727 \tabularnewline
\hline
Stress & Always-On & 0.9497 & 0.7486 & 0.0010 & 0.0000 & 0.0000 & 0.0000 \tabularnewline
Stress & Threshold-ES & 0.8580 & 0.7486 & 0.0010 & 0.0000 & 0.0000 & 0.2094 \tabularnewline
Stress & Threshold-ES-Hysteresis & 0.9497 & 0.7486 & 0.0010 & 0.0000 & 0.0000 & 0.0000 \tabularnewline
Stress & \textbf{Proposed-CRL-PPO} & 0.8500 & 0.7486 & 0.0010 & 0.0000 & 0.0000 & 0.1470 \tabularnewline
\hline
Unseen Stress & Always-On & 0.9213 & 0.6065 & 0.0010 & 0.0000 & 0.0000 & 0.0000 \tabularnewline
Unseen Stress & Threshold-ES & 0.6785 & 0.6065 & 0.0010 & 0.0000 & 0.0000 & 0.1671 \tabularnewline
Unseen Stress & Threshold-ES-Hysteresis & 0.7309 & 0.6065 & 0.0010 & 0.0000 & 0.0001 & 0.0098 \tabularnewline
Unseen Stress & \textbf{Proposed-CRL-PPO} & 0.7099 & 0.6065 & 0.0010 & 0.0000 & 0.0000 & 0.0722 \tabularnewline
\hline
\end{tabular*}
\end{table*}

Compared with Threshold-ES, the proposed method reduces switching activity under all scenarios while maintaining SLA-safe operation. Although threshold-based methods can save more energy in some cases, they are less adaptive and may either switch frequently or behave conservatively under stress traffic.

\subsection{Ablation Study}

\begin{table*}[t]
\centering
\caption{Ablation Study of Learning-Based Methods}
\label{tab:ablation_results}
\scriptsize
\setlength{\tabcolsep}{4pt}
\renewcommand{\arraystretch}{0.95}
\begin{tabular*}{\textwidth}{@{\extracolsep{\fill}}llcccccc@{}}
\hline
\textbf{Scenario} & \textbf{Method} & \textbf{Energy} & \textbf{Throughput} & \textbf{Delay} & \textbf{SLA Viol.} & \textbf{Thr. Loss} & \textbf{Switch Rate} \tabularnewline
\hline
Nominal & Unconstrained-PPO & 0.2905 & 0.3135 & 0.0059 & 0.8376 & 0.1213 & 0.1157 \tabularnewline
Nominal & Nominal-Only-PPO & 0.3706 & 0.2838 & 0.0010 & 0.0004 & 0.0015 & 0.0749 \tabularnewline
Nominal & Constrained-PPO-NoSwitch & 0.5154 & 0.2839 & 0.0011 & 0.0032 & 0.0019 & 0.1436 \tabularnewline
Nominal & \textbf{Proposed-CRL-PPO} & 0.5016 & 0.2840 & 0.0011 & 0.0072 & 0.0025 & 0.0727 \tabularnewline
\hline
Stress & Unconstrained-PPO & 0.8019 & 0.7831 & 0.0442 & 0.9992 & 0.1194 & 0.0213 \tabularnewline
Stress & Nominal-Only-PPO & 0.8726 & 0.7837 & 0.0097 & 0.4744 & 0.0297 & 0.2103 \tabularnewline
Stress & Constrained-PPO-NoSwitch & 0.8460 & 0.7486 & 0.0010 & 0.0000 & 0.0000 & 0.1513 \tabularnewline
Stress & \textbf{Proposed-CRL-PPO} & 0.8500 & 0.7486 & 0.0010 & 0.0000 & 0.0000 & 0.1470 \tabularnewline
\hline
Unseen & Unconstrained-PPO & 0.7188 & 0.6687 & 0.0360 & 0.9772 & 0.1529 & 0.0752 \tabularnewline
Unseen & Nominal-Only-PPO & 0.7490 & 0.6260 & 0.0042 & 0.4356 & 0.0472 & 0.3637 \tabularnewline
Unseen & Constrained-PPO-NoSwitch & 0.6922 & 0.6065 & 0.0010 & 0.0000 & 0.0000 & 0.0910 \tabularnewline
Unseen & \textbf{Proposed-CRL-PPO} & 0.7099 & 0.6065 & 0.0010 & 0.0000 & 0.0000 & 0.0722 \tabularnewline
\hline
\end{tabular*}
\end{table*}

Table~\ref{tab:ablation_results} compares the proposed method with learning-based ablations. Unconstrained-PPO is unsafe despite lower energy, and Nominal-Only-PPO fails under stress and unseen-stress traffic. In contrast, the constrained variants maintain zero SLA violation under stress and unseen-stress traffic, while the proposed method reduces switching compared with Constrained-PPO-NoSwitch.

\section{Reproducibility and Code Availability}

Code and experiment configurations are publicly available at https://github.com/dharmendra041983/sla-safe-energy-control-ngran.

\section{Conclusion}

The main contribution of this paper is a stability-aware constrained PPO framework for AI-native NG-RAN energy saving. The framework formulates energy control as a constrained closed-loop problem, combines adaptive Lagrangian penalties with mixed-regime training, and includes switching-cost regularization to reduce unnecessary energy-mode transitions. Evaluation in a seven-cell NG-RAN simulation shows that the proposed method achieves a balanced energy-QoS-stability tradeoff under nominal, stress, and unseen-stress traffic scenarios. The current evaluation relies on simplified traffic and channel assumptions; future work will evaluate the proposed controller under more realistic 5G NR channel conditions, including Rayleigh fading, dynamic uplink channel variation, and interference in dense urban deployments.

\end{document}